# Measurements of the intrinsic quantum efficiency and absorption length of tetraphenyl butadiene thin films in the vacuum ultraviolet regime

**Christopher Benson**[1,2,a], **Gabriel Orebi Gann**[1,2], **Victor Gehman**[2]

[1] University of California, Berkeley, CA 94720, USA
[2] Lawrence Berkeley National Laboratory, Berkeley, CA 94720, USA

**Abstract** A key enabling technology for many liquid noble gas (LNG) detectors is the use of the common wavelength shifting medium tetraphenyl butadiene (TPB). TPB thin films are used to shift ultraviolet scintillation light into the visible spectrum for detection and event reconstruction. Understanding the wavelength shifting efficiency and optical properties of these films are critical aspects in detector performance and modeling and hence in the ultimate physics sensitivity of such experiments. This article presents the first measurements of the room-temperature microphysical quantum efficiency for vacuum-deposited TPB thin films – a result that is independent of the optics of the TPB or substrate. Also presented are measurements of the absorption length in the vacuum ultraviolet regime, the secondary re-emission efficiency, and more precise results for the "black-box" efficiency across a broader spectrum of wavelengths than previous results. The low-wavelength sensitivity, in particular, would allow construction of LNG scintillator detectors with lighter elements (Ne, He) to target light mass WIMPs.

## 1 Introduction and motivation

Noble-gas detectors are becoming important to numerous experimental efforts involving: dark matter searches; neutrino and other particle detectors; searches for the neutron's electric dipole moment; and measurements of the neutron lifetime [1–15]. The properties of these elements, particularly in the liquid phase, are very attractive. They have exceptionally high scintillation yield (20,000–40,000 photons/MeV), which results in excellent energy resolution and low threshold. Their high density makes them more self-shielding than water or organic liquid scintillators. Tracking detectors can be constructed by applying an electric field across the bulk and collecting the ionization signal. The specifics of the scintillation process in noble gasses means they are almost transparent to their own scintillation light. Additionally, the time structure of that light is dependent on the incident particle type, allowing for pulse shape analysis to distinguish nuclear from electron recoils.

Detection of the vacuum ultraviolet (VUV) scintillation light produced by noble gas targets is a critical challenge common to this class of detectors. As shown in Fig. 1, the scintillation wavelengths can range from 175 nm for Xenon down to near 80 nm for Helium and Neon. Light of these wavelengths is strongly absorbed by most materials, including those commonly used for optical windows. Many experiments sidestep the issue of directly detecting VUV light though the use of wavelength shifting (WLS) films which absorb the VUV light and re-emit photons, typically in the visible spectrum. The visible photons can then easily be detected using photomultiplier tubes (PMTs).

As experimental programs look toward lower energies, the use of helium and neon as a detecting medium is attractive. The scintillation wavelengths of these noble gases are around 80 nm, deep into the VUV. This provides motivation to extend previous measurements of wavelength shifting materials into these extreme VUV regimes.

One commonly used WLS is tetraphenyl butadiene (TPB). TPB thin films have been widely studied [16–26] and are regularly used for many experimental programs [2–9] due to their relatively high efficiency. TPB may also be easily applied to surfaces using standard techniques, such as vacuum deposition in a thermal evaporator. The reemission spectrum of TPB has been measured and its shape well understood at cryogenic temperatures [22] and room temperature [16]. The wavelength shifting efficiency (WLSE) of thin films of TPB vacuum deposited on an acrylic substrate has been measured, without reference to other materials, in the range of

---

[a] e-mail: christopher.benson@berkeley.edu

Springer

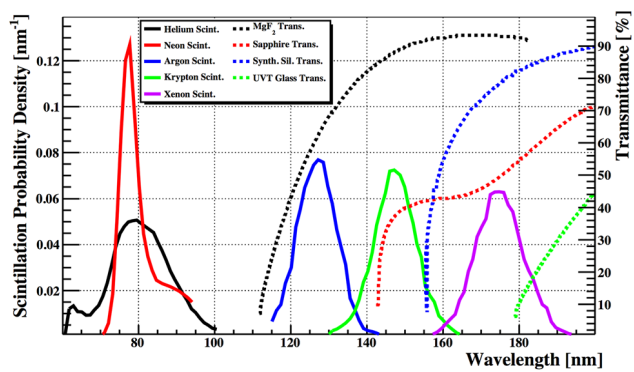

**Fig. 1** Scintillation wavelengths for various noble gases, along with the transmission of some commonly used optical windows [16]

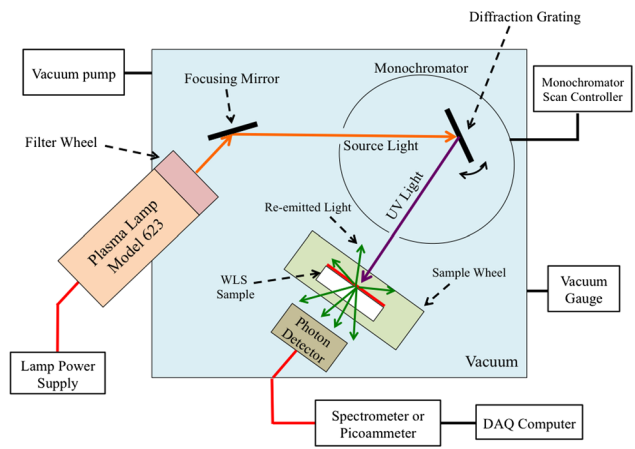

**Fig. 2** A cartoon schematic of the experimental setup for the long wavelength configuration (Sect. 3.1)

120–250 nm [16]. The WLSE as defined in [16] is a "black-box" definition of the efficiency that includes both the intrinsic QE of the TPB as well as certain optical properties of the TPB film and substrate. The resulting quantity is the efficiency that a photon absorbed by the sample is reemitted from the sample. This measurement is thus sample dependent, including effects of scattering and absorption, and cannot be directly applied to other apparatus.

This article presents the first ever measurement of the intrinsic quantum efficiency (QE) of TPB evaporated thin films, defined as the probability that a photon absorbed by a TPB molecule is reemitted, and its dependence on incident wavelength. Also presented are measurements of the absorption length of TPB in the VUV regime and a measurement of the average efficiency of the secondary reemssion process, referred to as SQE, in the region where the TPB absorption and reemission spectra overlap. As part of this work, room temperature measurements of the reemission spectrum of TPB as a function of incident wavelength are presented. The WLSE as defined in [16] is also studied, for the purpose of a direct comparison to this work. These results cover the spectral range studied by [16] with improved precision, while extending the measurements down to 50 nm.

Section 2 provides an overview of the hardware components of the experimental apparatus. Section 3 describes the various apparatus configurations used to cover the full wavelength region of interest (ROI). Section 4 describes the sample fabrication. Section 5 describes the data acquisition methods. Section 6 contains descriptions of the analysis methods used to calculate the absolute WLS efficiency, the detailed Monte Carlo model used to simulate the setup, and the techniques to extract the VUV QE, VUV absorption lengths and SQE. Section 7 presents results. Section 8 provides a comparison of this work to previous studies and discusses the impact of various improvements to the measurements. Conclusions are presented in Sect. 9.

## 2 Experimental apparatus

The primary objective of this work is to measure the intrinsic QE of TPB and the characteristic VUV photon absorption length in TPB, and its dependence on incident wavelength. Also measured are the room-temperature reemission spectrum and the black-box WLSE, with its dependence on sample thickness and incident wavelength. These measurements are performed by exposing TPB thin film samples to UV light of a known wavelength and intensity and measuring the amount and spectrum of light reemitted from fluorescence.

A cartoon representation of the experimental apparatus is shown in Fig. 2. The setup consists of a broad spectrum UV light source, filters, optical elements, a monochromator, a sample mounting assembly and photon detectors. All of the apparatus components are installed within a windowless vacuum chamber and held at vacuum. The setup is designed to produce and project monochromatic UV light onto thin film WLS samples in a repeatable fashion. A photodiode is used to separately measure the flux of photons incident on and reemitted by the samples. A spectrometer is used to measure the reemission spectrum of the samples.

Individual elements used in the experimental apparatus are described in this section. Because this work covers a wide range of wavelengths (50–250 nm), several distinct hardware configurations are used. Each of these configurations are constructed from elements described in this section. The distinct apparatus configurations are described in Sect. 3.

### 2.1 UV light source

This work uses two distinct light sources. Each light source is specialized to a subset of the entire 50–250 nm wavelength ROI. Only one light source may be attached to the setup at a time.

### 2.1.1 Deuterium light source

The McPherson Model 623 light source, referred to as the Deuterium Light Source (DLS), produces light for measurements in the range of 125–250 nm. The DLS contains a sealed deuterium gas volume which is ionized to produce broad spectrum light with a bright UV component. The output light passes through a circular 1" diameter $MgF_2$ exit window. The $MgF_2$ window has a transmission cutoff wavelength of 115 nm which drives the lower limit of emission for this source [27]. The DLS is powered using a McPherson Model 732 power supply.

### 2.1.2 Windowless light source

The McPherson Model 629 Hollow Cathode Gas Discharge Source, referred to as the windowless light source, produces light for measurements in the range of 45–150 nm. Unlike the DLS, the windowless light source does not have a window at its output. A feed gas is supplied to the lamp at a constant pressure and is ionized using high voltage. The spectrum of light output by the lamp depends on the choice of the feed gas. Because there is no sealing window between the light source and the rest of the setup, a differential pumping port near the lamp output is used to reduce the gas load on the vacuum from the feed gas flowing out of the light source's output.

## 2.2 Optical chain

This section describes the elements that make up the optical chain. These are the filter wheel, focusing mirror and monochromator.

### 2.2.1 Filter wheel

An Action Research Corporation Model 52 filter wheel is coupled directly downstream of the light source. The filter wheel is used to apply optical filters to the broad spectrum light output from the light source for certain classes of measurements (Sect. 5). The filter wheel has four slots which can each hold a 1.90-cm diameter disk. Changing the selected filter slot is done using an exterior nob and is possible while the setup is under vacuum. When a slot is selected, the filter is placed concentrically in the light output from the light source. All light must pass through the selected filter wheel slot to continue along the optical chain.

Three of the four slots in the filter wheel are used. The first slot is empty, which allows for a no-filter condition. The second slot contains a 0.48-cm thick, uncoated fused quartz silica filter which has a 155 nm cutoff. The third slot holds a 0.48-cm thick, uncoated $MgF_2$ with a 115 nm cutoff. The fused quartz silica and $MgF_2$ filters are used as high-pass optical filters which allow measurements of background levels (Sect. 5.1.4).

### 2.2.2 Focusing mirror

A McPherson Model 615 focusing elbow sits between the filter wheel and monochromator entrance slit. The focusing elbow contains a curved $Al + MgF_2$ focusing mirror which focuses the light passing through the filter wheel onto the monochromator entrance slit. The focus of the mirror is adjusted using three set screws and is configured to maximize the light output at the monochromator's exit for wavelengths in the ROI.

### 2.2.3 Monochromator

A McPherson Model 234/302 Vacuum Ultraviolet Monochromator is used to output monochromatic light of a selected wavelength from a broad spectrum input. The light entering the monochromator passes through an entrance slit which is 1.78-mm wide and 4.88-mm high. The entrance slit projects incoming light onto a rotatable holographic diffraction grating positioned near the center of the monochromator. The angle between the diffraction grating and the incoming light determines the wavelength of light projected from the grating onto the monochromator's exit slit. The diffraction grating's angular position is adjusted and set using a servo motor which is controlled using the McPherson Model 789A-3 Scan Controller. The exit slit of the monochromator has the same dimensions as the entrance slit. Figure 3 shows a comparison of monochromator wavelength setting and the measured spectrum of the output light. The peak of the output light spectrum is in good agreement with the monochromator setting.

This work uses two types of diffraction gratings. The monochromator only holds one grating at a time. Similar to the light sources, each diffraction grating is specialized to a subset of wavelengths in the ROI. The two diffraction gratings are an $Al + MgF_2$ coated grating with 1200 gratings per mm, and a Platinum (Pt) coated grating with 2400 gratings per mm. As shown in Fig. 4, the $Al + MgF_2$ coated grating outperforms the Pt coated grating for wavelengths greater than 105 nm.

## 2.3 Sample holder

The monochromator's outlet attaches to a McPherson Model 648 Vacuum Filter Wheel, referred to as the sample wheel. The sample wheel contains five 2.54-cm diameter slots for sample disks. The sample disks are installed in the wheel and fixed in place using snap rings. The selected slot in the sample wheel is changed using an external knob and may be adjusted when the setup is under vacuum. The adjustment knob allows the cycling of multiple samples in and out of the UV beam during a data acquisition run. When a slot is selected, the sample slot is placed concentrically in the monochromatic UV beam exiting the monochromator. The sample slots are labeled 1–5. Slot 1 contains a thick aluminum disk, used to perform a dark current measurement. Slot 2 is left empty to allow for a measurement of the total VUV flux incident on the samples from the monochromator's output. Slots 3 through 5 hold WLS samples to be studied.

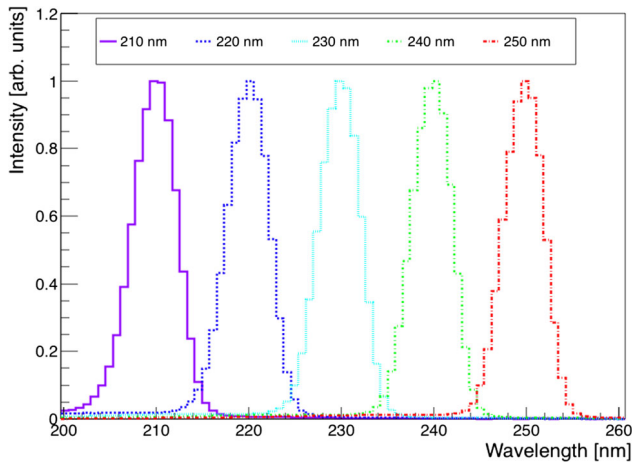

**Fig. 3** A set of light specta measured downstream of the monochromator's exit slit for several wavelength settings using an Al + MgF$_2$ diffraction grating. Each label indicates what the monochromator was set to when each spectrum was taken. There is good agreement between the monochromator setting and the output spectrum. All spectra peaks are normalized to one

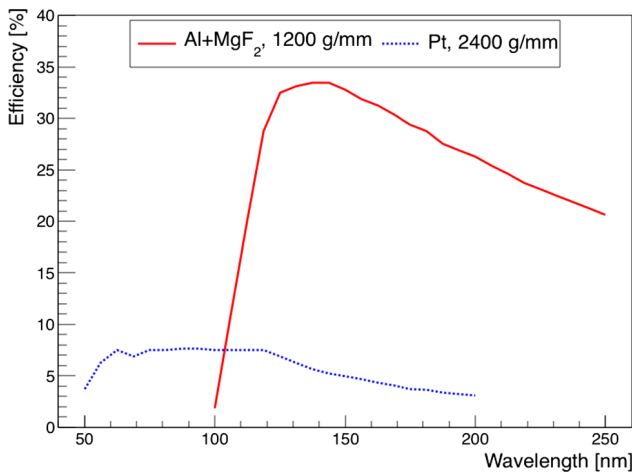

**Fig. 4** A comparison of the reflective efficiecnies for an Al + MgF$_2$ coated grating and a Platium (Pt) coating grating. Plot data provided by McPherson Inc. [28]

### 2.4 Photon detection

This work uses two types of photon detectors: a photodiode and a spectrometer. The setup uses one of these detectors at

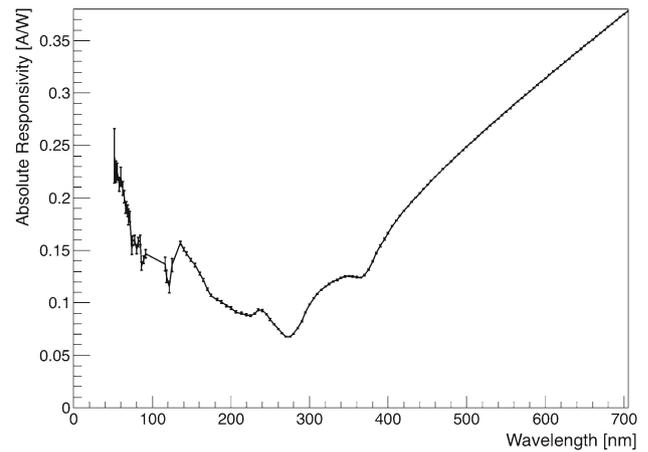

**Fig. 5** The AXUV100G photodiode [29] absolute responsivity calibration as a function of wavelength from NIST in the range of 50–700 nm. The photodiode was calibrated in February 2016 [32] for wavelengths 50–1100 nm

a time. As shown in Fig. 2, photon detection occurs downstream of the sample wheel.

#### 2.4.1 Photodiode

An Opto Diode AXUV100G photodiode [29–32] is used to measure the flux of VUV light incident on samples as well as the flux of reemitted light from WLS samples in the sample wheel. The device is a passive, windowless photodiode cell with an active area of 1 cm by 1 cm. A vacuum electrical feedthrough and coaxial cable (outside of the vacuum space) electrically couple the photodiode to an external Keithley 485 picoammeter for photocurrent readout. A data acquisition computer connected to the picoammeter queries real-time current readings using a LabVIEW graphical user interface (GUI) and stores the values for offline analysis. The absolute response of the photodiode was calibrated by National Institute of Standards and Technology (NIST) in February of 2016 [32] in the range of 50–1100 nm. The absolute response as a function of wavelength is shown in Fig. 5.

#### 2.4.2 Spectrometer

An Ocean Optics QE65000 spectrometer is used to measure the reemission spectrum of WLS samples. A vacuum feedthrough assembly consisting of a collimating lens and fiber optic allows light to be collected and routed out of the vacuum space for analysis. A 200-$\mu$m diameter quartz fiber couples the vacuum feedthrough assembly output to an input port on the spectrometer.

The spectrometer is sensitive to wavelengths in the range 200–1000 nm. The Ocean Optic's Spectra Suite software package installed on the data acquisition computer is used to configure and read out the spectrometer.

## 3 Apparatus configurations

Three apparatus configurations are used to cover the ROI (50–250 nm). These are the long wavelength, intermediate wavelength, and short wavelength configurations. Each configuration specializes to a subset of VUV wavelengths in the ROI and is composed of the hardware elements described in Sect. 2. The long wavelength and intermediate wavelength configurations overlap in the range of 130–150 nm which allows for cross checks of configuration–dependent systematic uncertainties.

The hardware composition of each configuration maximizes the intensity of VUV light exiting the monochromator within its particular wavelength range. Maximizing VUV intensity is done to minimize statistical uncertainties in measurements.

### 3.1 Long wavelength configuration

The Long Wavelength Configuration (LWC) is used to measure the WLSE and reemission spectra of samples for incident wavelengths in the range of 125–250 nm. As shown in Fig. 2, this configuration uses the DLS, filter wheel, focusing mirror, monochromator with the $Al + MgF_2$ diffraction grating and sample wheel.

### 3.2 Intermediate wavelength configuration

The Intermediate Wavelength Configuration (IWC) is used to measure the WLSE and reemission spectra of samples for incident wavelengths in the range of 100–150 nm. A cartoon representation of this configuration is shown in Fig. 6. This configuration uses the windowless light source, filter wheel, monochromator with the $Al+MgF_2$ grating and sample wheel. Unlike the LWC, this configuration uses the windowless light source and does not use the focusing mirror. A high purity $N_2$ or argon feed gas is used in the windowless light source in this configuration.

### 3.3 Short wavelength configuration

The Short Wavelength Configuration (SWC) is used to measure the WLSE and reemission spectra of samples for incident wavelengths in the range of 45–100 nm. This configuration is very similar to the IWC. The only differences are that a Pt diffraction grating is used instead of the $Al + MgF_2$ grating, and the feed gas used for the windowless light source is a 90% Neon, 10% Helium mixture.

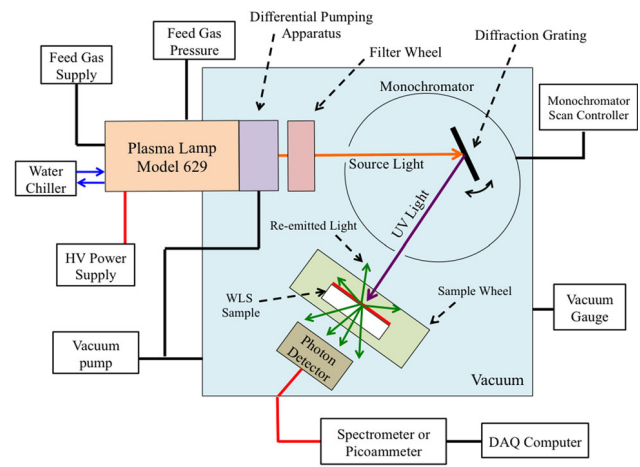

**Fig. 6** A cartoon schematic of the experimental setup for the IWC and SWC (Sects. 3.2 and 3.3)

**Table 1** Summary of samples fabricated and studied in this work. The thickness column referes to the TPB film thickness as measured by a profilometer

| Sample | Thickness ($\mu$m) | Uncert. ($\mu$m) |
|---|---|---|
| A | 0.5 | 0.1 |
| B | 0.7 | 0.1 |
| C | 0.9 | 0.1 |
| D | 1.8 | 0.2 |
| E | 2.2 | 0.2 |
| F | 2.55 | 0.2 |
| G | 3.1 | 0.2 |
| H | 3.7 | 0.2 |

## 4 Sample fabrication

Several TPB thin-film samples were fabricated using thermal evaporators at the Molecular Foundry at Lawrence Berkeley National Laboratory and in the Dr. Daniel McKinsey Laboratory at the University of California at Berkeley. A tabulation of samples used in this work are shown in Table 1. The films were deposited on stabilized ultraviolet-transmitting (SUVT) acrylic manufactured by Polymer Plastics Company, LC. A stock acylic sheet was cut into circular disks with a 2.54-cm diameter and 0.318-cm thickness. The transmission of the SUVT acrylic, shown in Fig. 8, was measured and shown to be consistent with the manufacturer's data sheet [33]. The acrylic's transmission is 90% for wavelengths longer than 300-nm and 0% for wavelengths below 250-nm, as shown in Fig. 8. This is important because the acrylic substrate should behave as a high-pass filter, being transparent to the reemitted light from the TPB but opaque to the UV light incident on the sample.

Evaporative deposition was performed at a pressure less than 5E-6 torr. Film thicknesses between 0.5 $\mu$m and 3.7 $\mu$m were chosen so that our samples would be comparable in thickness to those used in [16] and in the MiniCLEAN and DEAP-3600 experiment.

The final thicknesses of the TPB thin films were measured using a Dektak Profilometer. Before evaporative deposition, two small pieces of Kapton tape were placed on opposite edges of the acrylic substrate. Following evaporation, the Kapton tape was removed to provide a location to measure film thickness. This was done by measuring the height of the step between the acrylic substrate (where the Kapton was attached) and the TPB film. Step measurements were preformed at least three times at each location and averaged. Measurements at opposite locations of the sample were within the quoted uncertainty, indicating that thickness gradients across the sample were small. No attempts to directly measure the gradient profile across the full surface of the sample were performed due to a concern that the profilometer tip could damage the TPB surface. As conservative measure, the quoted uncertainty in TPB film thickness (Table 1) is larger than the any difference in film thickness observed at opposite ends of the samples.

Due to evidence that ambient ultraviolet light can degrade the WLS performance of TPB thin films [20,21], after fabrication the samples were stored in a separate dark, clean vacuum chamber held at a pressure of less than 1E-1 torr using an oil-less diaphragm pump. Each sample was measured within a month of its fabrication, and each sample was only measured up to three times in each configuration. Results from the repeated measurements were consistent with each other, suggesting no degradation for this level of exposure to within the uncertainty of the measurement.

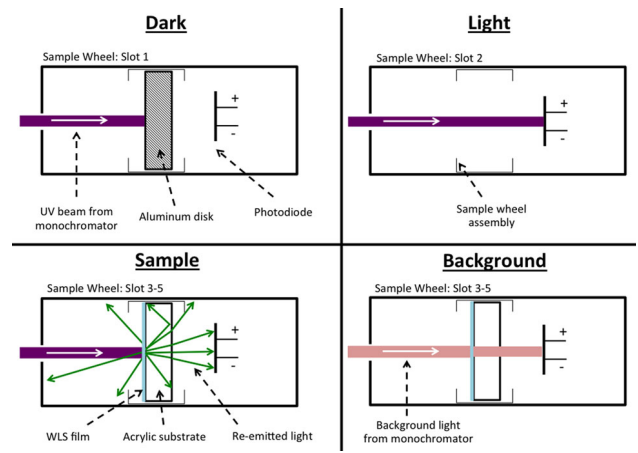

Fig. 7 A cartoon representation of the four types of photocurrent measurements

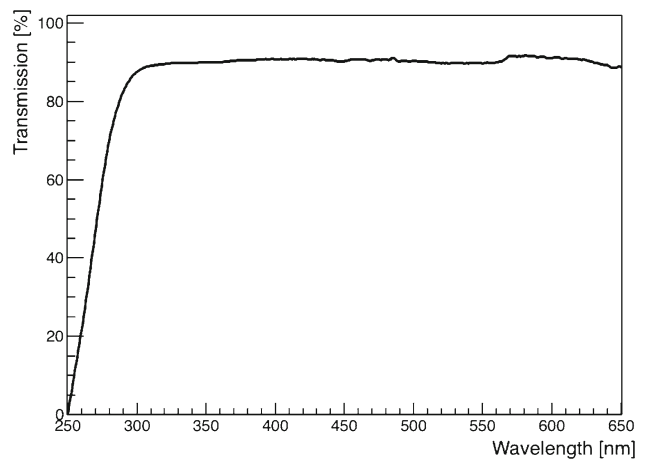

Fig. 8 Transmission as a function of wavelength for 2.54-cm diameter and 0.32-cm thick SUVT acrylic disk. This was measured using the spectrometer and the LWC

## 5 Measurement procedure

This section describes the two classes of measurements performed in this work: photocurrent readout for WLSE measurements, and reemission spectra.

### 5.1 Photocurrent measurements

A measurement of the photocurrent output by the photodiode is required to determine the flux of VUV light incident on the samples as well as the flux of reemitted light captured by the photodiode. These photocurrent measurements are used as inputs in the WLSE calculations, described in Sect. 6. The four types of photocurrent measurements made are Dark, Light, Sample and Background. Each of these are described in this section, and are illustrated in Fig. 7.

#### 5.1.1 Dark photocurrent

The dark photocurrent measurement provides a baseline measurement of the dark current for subtraction. This is shown in the top left of Fig. 7. It is performed by selecting slot 1 in the sample wheel, which contains a 2.54-cm diameter aluminum disk with 0.635-cm thickness. The aluminum disk blocks any light from reaching the photodiode, allowing for a proper baseline measurement.

#### 5.1.2 Light photocurrent

The light photocurrent measurement provides a measurement of the flux of VUV light incident on the samples. This is done by selecting slot 2 on the sample wheel, which is empty. A representation of this is shown in the top right portion of Fig. 7. The photodiode active area fully captures all of the UV light that would be absorbed by a sample placed in the beam.

### 5.1.3 Sample photocurrent

Theسample photocurrent measurement provides a measurement of the reemitted light flux at the photodiode's location. This measurement is performed on slots that contain WLS samples (slots 3–5). A representation of this measurement is shown in the bottom left of Fig 7. The UV beam from the exit slit of the monochromator is absorbed by the WLS thin film and the acrylic substrate. Photons are reemitted from the WLS film in the forward and backward direction according to a currently unknown angular distribution. A portion of the total number of reemitted photons are collected by the photodiode. The determination of the total amount of reemitted light from this measurement is described in Sect. 6.1.

The sample photocurrent measurements are performed less than one minute after the light photocurrent measurement. The lamp output intensity was confirmed to be stable on timescales much longer than one minute thus providing confidence that the flux of incident VUV light on the sample does not change between a light photocurrent measurement and a sample photocurrent measurement.

### 5.1.4 Background photocurrent

The background photocurrent measurement provides a measure of the stray light contamination in the UV beam and its contribution to the light and sample photocurrent measurements. This measurement is performed by applying a high-pass optical filter using the filter wheel located near the light source (Sect. 2.2.1). The filter absorbs the component of the light source's output below the cutoff wavelength of the filter. For monochromator wavelength settings below the cutoff wavelength of the selected filter, the UV light normally exiting the monochromator, which is used to drive the reemission from WLS samples, is eliminated. This leaves any background light above the filter's cutoff incident on the sample. A sample is left in the beam during the measurement to account for the transmission of the thin film. The possibility of fluorescence originating from the filter was crosschecked by measuring the light flux with no sample in the beam, and with a filter applied for an incident VUV wavelength below the cutoff of the filter. The light flux was found to be consistent with background for both filters at all incident VUV wavelengths of interest below the cutoff wavelength of the filters. This verifies that any fluorescence originating from the filters is negligible to within our sensitivity.

The possibility of fluorescence originating from VUV photons terminating in the acrylic substrate was crosschecked by performing a sample photocurrent measurement on a blank acrylic disk. This photocurrent measurement was found to be consistent with background which verifies that any fluorescence of the acrylic from VUV photons is below our sensitivity.

It should be noted that light and sample photocurrent measurements are often several orders of magnitude greater than the background photocurrent; background current corrections are negligible for wavelengths in the range 100–250 nm. The background corrections become important when the sample photocurrent is of the order of the background current. For wavelengths below 100 nm (the measurements performed at 50 and 73 nm) the correction is non-negligible; at its peak, the background can reach roughly one half of the sample photocurrent. In all cases, the uncertainty on the background is included in the final photocurrent measurement.

## 5.2 Spectrometer measurements

Reemission spectra are measured using the setup described in this section. These spectra are used as inputs for the efficiency calculations discussed in Sect. 6.

The spectrometer is configured to integrate for 10 s for each spectral measurement. Three spectral measurements are averaged online and the result is written to disk as a text file. This is repeated 20 times during the course of a data run resulting in 20 files written to disk. The files written to disk are used as inputs for an offline analysis.

Two classes of data runs are required to measure a WLS sample's reemission spectrum: Dark and Sample. In the same way as the dark photocurrent measurement, a dark spectrum is measured by selecting the aluminum disk in slot 1 of the sample wheel to block all light from reaching the spectrometer. This provides a baseline for subtraction in an offline analysis. The sample spectrum is measured by selecting the appropriate slot on the sample wheel (slots 3–5). This is exactly analogous to the sample photocurrent measurement shown in Fig. 7, because only the spectrum of reemitted light is measured in this configuration.

An offline analysis is performed using the ROOT data analysis package [34] to extract the corrected reemission spectrum. The measurements from the dark spectrum run are averaged and subtracted from the average of the WLS sample reemission spectrum run. The result is then corrected by the acrylic transmission (Fig. 8) and the relative transmittance of the collimating lens/fiber assembly and quartz fiber (Fig. 9).

## 6 Analysis methods

The absolute "black-box" WLSE of a fluorescing thin film for incident light of wavelength $\lambda$ is defined to be the ratio of the number of photons reemitted by the sample (film and substrate) to number of photons incident on the film. Equivalently, this can be interpreted as the probability that an incident photon of a certain wavelength will be absorbed and reemitted, at a wavelength according to the reemission spectrum, and escape the WLS sample.

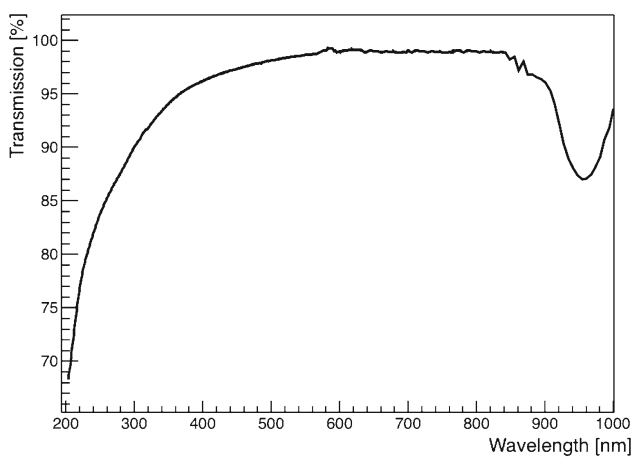

**Fig. 9** Relative transmittance of collimating lens/fiber vacuum feedthough and 200 $\mu$m fiber leading to the spectrometer

As discussed in Sect. 1, the WLSE defined and measured in this way combines the true, microphysical QE of the material with optical effects of the film and the substrate, treating the sample as a "black box". It is the efficiency with which the black-box sample wavelength shifts incident photons, rather than the efficiency for a TPB molecule to do so. Because reemitted photons can be reabsorbed before they escape the film, the WLSE defined in this was has a dependence on film thickness and is a property of the specific sample, rather than a property of the material. Physically, the underlying QE is the more interesting result because it is an intrinsic property which should not depend on film thickness.

The VUV QE, VUV absorption length and SQE of TPB are determined by measuring the WLSE of several films of different thickness and unfolding the intrinsic microphysical parameters from the optical effects of the thin film by comparing to a detailed microphysical simulation. Section 6.1 describes the method of calculating the absolute WLSE of a sample from raw photocurrent data. Section 6.2 describes the details of the Monte Carlo model and Sect. 6.3 discusses the method to extract the microphysical parameters from measurements using the model.

### 6.1 Absolute wavelength shifting efficiency

Because the WLSE of a fluorescing thin film is defined as a ratio of photon fluxes, the total flux of photons incident on and reemitted by a wavelength-shifting sample must be determined.

The total flux of UV photons incident on a sample is determined from the light photocurrent measurements (Sect. 5.1.2). The geometry of the setup is arranged such that all of the UV light incident on the samples is captured by the photodiode, which allows for a direct measurement of the total flux of UV photons incident on the sample.

The flux of reemitted photons detected at the location of the photodiode is determined from the sample photocurrent measurement (Sect. 5.1.3). Because the photons reemitted by the TPB are emitted in the forward and backward directions and according to a currently unknown but usually assumed Lambertian angular distribution, only a fraction of the total reemitted photons are collected by the photodiode during a sample photocurrent measurement. The ratio of reemitted photons collected by the photodiode to the total number of reemitted photons leaving the sample is defined as the geometric acceptance fraction (GAF).

The GAF is interpreted as the probability that a reemitted photon that has escaped the sample (TPB film + acrylic disk) is observed by the photodiode. The GAF is independent of the choice of modeled QE because the GAF is simply a ratio of the number of visible photons detected to the number of visible photons escaping the TPB sample – it represents the geometric acceptance of the photodiode. Changing the VUV QE parameter changes the normalization (the average number of visible photons that are produced) but does not change the fraction of these that are detected. This expected behavior was verified with the Monte Carlo simulation. The GAF is dependent on specifics of the setup's geometry, with a second-order dependence on the VUV absorption length and TPB thickness. The latter arises due to re-absorption and scattering of visible photons in the bulk TPB, which alters the angular distribution of photons emitted from the sample and, thus, the fraction observed by the photodiode.

The GAF is determined from a detailed microphysical Monte Carlo (Sect. 6.2) simulation of the setup and is used to determine the total reemitted photon flux from the measured reemitted photon flux.

The measured photocurrents are a convolution of the spectrum of light incident on the photodiode, multiplied by the photon energy, with the photodiode's calibrated response (Fig. 5), $R(\lambda)$. For light photocurrent measurements at wavelength $\lambda$, the incident light spectrum is given by the wavelength distribution at the monochromator's exit, $M(\lambda - \lambda')$, centered around $\lambda$. As shown in Fig. 3, $M(\lambda - \lambda')$ can be accurately modeled as a Gaussian with the width set by the type of diffraction grating used in the monochromator. For sample photocurrent measurements the TPB reemission spectrum, $P(\lambda)$, is used. The reemission spectrum of TPB was shown to be constant for illumination wavelengths from 128–250 nm in [16]. This has been verified in this work and has been extended down to 45 nm incident light (Sect. 7.1).

The light photocurrent, $I_{\text{light}}(\lambda)$, and sample photocurrent, $I_{\text{TPB}}(\lambda)$, measurements are corrected for dark photocurrent (Sect. 5.1.1) and background photocurrent (Sect. 5.1.4). The dark photocurrent measurement provides a baseline correction while the background photocurrent corrects for stray light components in $M(\lambda - \lambda')$. The total photocurrent correction, $I_{\text{corr}}(\lambda)$, is the sum of the dark photocurrent and background contributions and is defined in Eq. 1.

$$I_{\text{corr}}(\lambda) = \begin{cases} I_{\text{dark}} + I_{\text{background}} & \text{if } \lambda \leq 150 \text{ nm} \\ I_{\text{dark}} & \text{if } \lambda > 150 \text{ nm} \end{cases} \quad (1)$$

The background photocurrent is included in $I_{\text{corr}}(\lambda)$ for wavelengths less than or equal to 150 nm instead of the full ROI due to the filter availability. As described in Sect. 5.1.4, background measurements can only be performed below the cutoff wavelength of the filter. The longest available cutoff wavelength is the fused quartz scilica filter with a cutoff wavelength of 150 nm. It should be noted that only measurements using the LWC (Sect. 3.1) for wavelengths above 150 nm do not include background corrections. It was verified using the spectrometer and photocurrent measurements on uncoated acrylic disks that background levels in this range are consistent with dark current measurements i.e. below our sensitivity, thus eliminating the need for background measurements in this range.

As discussed in Sect. 6, the WLSE is computed by taking the ratio of the flux of reemitted light at the photodiode to the flux of incident light and dividing by the GAF, $A_i$. For convenience, the ratio of the measured photon fluxes, $\beta_{i,exp}(\lambda)$, and the WLSE , $\varepsilon_i(\lambda)$, are defined separately. These values are calculated for each sample where $i$ denotes the sample index. The measured photon ratio is given in Eq. 2:

$$\beta_{i,exp}(\lambda) = \frac{I_{\text{TPB}}(\lambda) - I_{\text{corr}}(\lambda)}{I_{\text{light}}(\lambda) - I_{\text{corr}}(\lambda)} \times \frac{\int d\lambda' \frac{hc}{\lambda'} R(\lambda') M(\lambda - \lambda')}{\int d\lambda'' \frac{hc}{\lambda''} R(\lambda'') P(\lambda'')}. \quad (2)$$

The WLSE of the ith sample is given in Eq. 3:

$$\varepsilon_i(\lambda) = \beta_{i,exp}(\lambda) \times \frac{1}{A_i}. \quad (3)$$

### 6.2 Monte Carlo simulation

The purpose of the Monte Carlo simulation is two fold. First it is used to determine the GAF for calculation of the absolute WLSE (Sect. 6.1), and second, to unfold several underlying microphysical parameters of TPB from the optical effects (Sect. 6.3).

The simulation is performed using the Reactor Analysis Tool (RAT). RAT is a detailed GEANT4 based, microphysical simulation and analysis framework first written for the Braidwood experiment [35]. Versions of RAT are currently being used by several experiments, including SNO+ [36], DEAP-3600 [6] and MiniCLEAN [7]. This work uses the MiniCLEAN version of RAT, which contains the functionality to simulate photon absorption and reemission of TPB thin films.

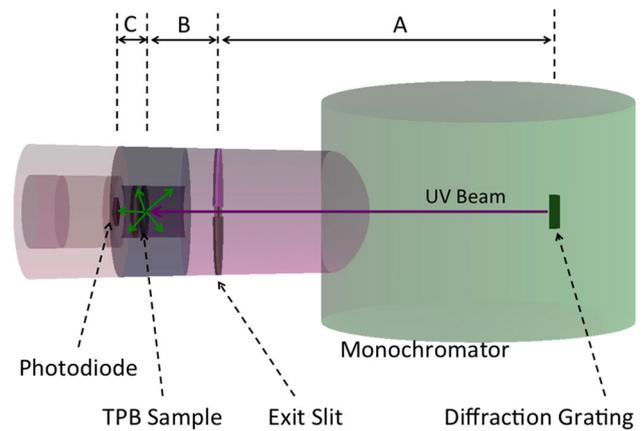

**Fig. 10** A rending of the experimental setup as modeled in RAT. Several key elements and dimensions are labeled. Values and uncertainties for the dimensions are given in Table 2. Dimension A is the distance from the center of the diffraction grating face to the monochromator exit slit. Dimension B is the distance from the exit slit to the TPB surface. Dimension C is the distance from the TPB surface to the photodiode surface

**Table 2** Important apparatus dimensions and their uncertainties

| Dimension | Value (cm) | Uncert. (+/−) (cm) |
|---|---|---|
| A | 18.33 | 0.07 |
| B | 4.003 | 0.025 |
| C | 0.579 | 0.016 |
| Exit slit height | 0.152 | 0.013 |
| Exit slit width | 0.488 | 0.005 |
| Acrylic thickness | 0.318 | 0.013 |

#### 6.2.1 Model

The dimensions of the experimental apparatus were carefully measured and a model was constructed in RAT. Each dimension was measured 5 times. The average of the 5 measurements was used while the RMS provided the uncertainty. A rendering of the geometry as modeled in RAT is shown in Fig. 10. Important dimensions are provided in Table 2.

In the simulation, monochromatic photons are generated at the center of the diffraction grating surface. The photons are generated one at time and in the direction of the monochromator's exit slit. The angular distribution of the photons leaving the source vertex uniformly fills a cone with a 2° opening angle.

Many of the generated VUV photons terminate on the monochromator's exit slit. The fraction that pass through the slit propagate to the TPB surface boundary where they pass into bulk TPB or are reflected, according to Snell's law. The index of refraction for bulk TPB was estimated by a member of the MiniCLEAN collaboration [37] using the methods in [38] and compared to values of similar molecules in [39]. A value of $1.67 \pm 0.05$ is used in the simuation for all wavelengths. The TPB/vacuum interface is modeled as a rough surface using the GLISUR surface model [40] in GEANT4 with a polish value of $0.01 \pm_{0.01}^{0.09}$.

VUV photons that enter the bulk TPB are then absorbed according to wavelength–dependent absorption lengths. The absorption lengths of photons with wavelengths less than 250 nm in TPB were determined by fitting to the thickness dependence of the photon ratio for each incident wavelength in the ROI, as described in Sect. 6.3. Photons are reemitted isotropically at the same vertex where the VUV photon was absorbed and according to the measured visible reemission distribution, which is consistent with [16]. The average number of photons reemitted when a VUV photon is absorbed is determined by a chosen value of the QE. The simulation assumes energy must be conserved, meaning the sum of reemitted photon energies is less than or equal to the absorbed VUV photon energy.

The absorption length for photons in TPB for wavelengths greater than 250 nm are taken from [19] and is shown in Fig. 11. The TPB reemission and absorption spectra overlap in the blue tail, thus it is possible for photons to be absorbed and reemitted several times. This secondary reemission leads to red-shifting of the TPB reemission spectra in thicker samples [22]. This effect is naturally included in the Monte Carlo model. The efficiency of this secondary reemission process, referred to as secondary QE (SQE), is evaluated from our data. The SQE is interpreted as the average intrinsic QE of TPB in the wavelength region where the absorption and reemission spectra overlap. Determination of the SQE is discussed in Sect. 6.3.

Diffuse scattering of visible light occurs in TPB. To account for this, the Rayleigh scattering of visible reemitted photons is modeled. A mean scattering length of $2.75 \pm_{0.85}^{0.15}$ μm is used for reemitted photons [23]. Reemitted photons are produced isotropically at the time of creation. Scattering within the TPB layer is observed to produce the expected behavior of an approximate Lambertian angular distribution in the forward/backward directions for photons exiting the TPB sample.

The reemitted visible photons produced in the TPB are propagated until they are reabsorbed by the TPB or absorbed on walls, the surface of the photodiode, or by the acrylic. All photons which terminate on the photodiode surface are assumed to be detected. Optical effects, such as refraction and reflections, are included in the simulation. The measured

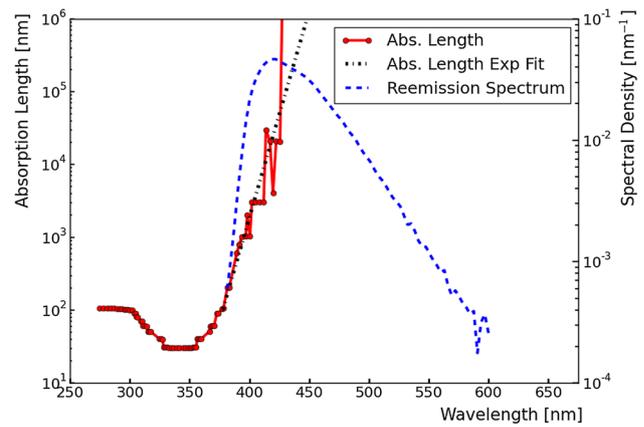

**Fig. 11** Absorption length of photons in TPB in nanometers plotted with the area normalized reemission spectrum of TPB. The absorption was taken from [19] while the reemission spectrum was measured in this work

SUVT acrylic transmission (Fig. 8) and manufacturer supplied index of refraction are used in the simulation of the acrylic. The tracks of all photons are stored in a ROOT file for post simulation analysis.

A python analysis script loops over the stored photon tracks and calculates values of interest. For comparison to data, the photon ratio, $\beta_{i,sim}$, is evaluated by taking the ratio of the number reemitted photons which terminate on the photodiode surface, $N_{pd}$, to the number of VUV photons which were incident on the TPB surface, $N_{inc}$, in the simulation. As expected, $\beta_{i,sim}$ depends on the choice of the QE, the sample thickness, and the optical properties of TPB and other materials which are constrained by measurements and values from the literature. $\beta_{i,sim}$ is defined in Eq. 4 as:

$$\beta_{i,sim}(QE) = \frac{N_{pd}}{N_{inc}}. \quad (4)$$

### 6.3 Microphysical parameter extraction

As discussed at the beginning of Sect. 6, the VUV QE, VUV absorption length, and SQE of TPB for a given wavelength of incident VUV light is determined by comparing the thickness-dependent response of samples to a detailed microphysical Monte Carlo simulation. By comparing the photon ratio measurements to simulation, it is possible to unfold the intrinsic microphysical properties from the optical effects of the sample and experimental setup.

More specifically, the measured photon ratio, $\beta_{i,exp}$, is compared to the photon ratio from simulation, $\beta_{i,sim}$, for samples of different thickness. For a given set of model value inputs for the TPB VUV QE, VUV absorption length, and SQE, the $\chi^2$ difference between data and simulation is evaluated. The $\chi^2$ is defined in Eq. 5 as:

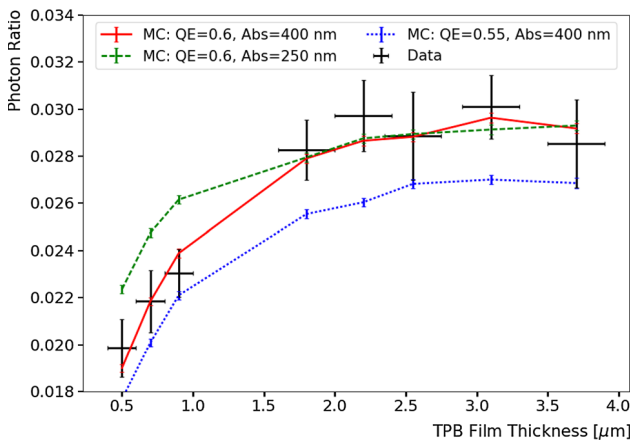

**Fig. 12** The measured, $\beta_{i,exp}$, and simulated, $\beta_{i,sim}$, photon ratios plotted as a function of sample thickness for data and Monte Carlo for several choices of VUV QE and VUV absorption length for 130-nm incident light. The SQE is held fixed at 0.9

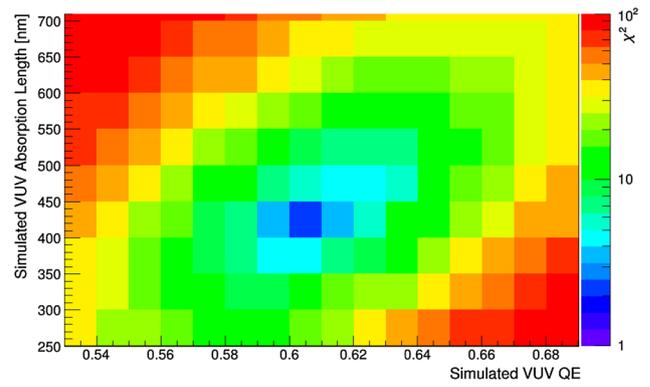

**Fig. 13** A heat map of the $\chi^2$ for a 2-D sweep of VUV QE and absorption lengths for 130 nm incident light. The SQE is held fixed at 0.9. The pair of VUV QE and absorption length values that minimizes the $\chi^2$ is 0.6 and 400 nm. The $\chi^2$ value at the minimum is equal to 2.0

$$\chi^2 = \sum_{i=1}^{n} \left( \frac{\beta_{i,exp} - \beta_{i,sim}}{\sigma_{i,exp}} \right)^2. \qquad (5)$$

The preferred value for each of these microphysical parameters is found by minimizing the $\chi^2$. Figure 12 shows the measured photon ratio along with several simulated photon ratios using different values of VUV QE and VUV absorption length. The solid line represents VUV QE and VUV absorption length choices that minimize the $\chi^2$ for the case of 130-nm incident light, while the dotted and dashed lines show the effect of sub-optimal choices for these parameters. As seen by the dotted line, changing the VUV QE affects the overall amount of light produced, which scales the photon ratio up and down. The dashed line illustrates that adjusting the VUV absorption length affects the shape of the thin-sample tail while leaving thicker samples unchanged. This behavior is expected because when the VUV absorption length is on the order of the sample thickness a non-negligible fraction of VUV photons will pass through the TPB and be "lost" in the acrylic substrate. As the simulated VUV absorption length is decreased the effect is thus more significant for thinner samples: more VUV photons start to be absorbed in these samples thus increasing the extracted response, or simulated photon ratio. The effect on thicker samples is small because close to 100% of the VUV photons penetrating the film are absorbed in either case.

Figure 13 shows a heat map of $\chi^2$ values for many choices of VUV QE and absorption length for 130 nm incident light. A minimum $\chi^2$ can be seen clearly at values of 0.6 and 400 nm for the VUV QE and absorption length respectively. This 2-D parameter sweep, as shown in Fig. 13, is performed for each incident wavelength studied in the ROI to identify the preferred VUV QE and absorption length values for a fixed value of the SQE.

As discussed in Sect. 6.2.1 and seen in Fig. 11, the TPB absorption spectrum overlaps with the blue tail of the TPB reemission spectrum. This allows for secondary reemission, where blue reemitted photons may be reabsorbed and reemitted by the TPB film. The efficiency of this process is referred to as the secondary reemission QE (SQE).

To evaluate the preferred value of the SQE, a 3-D scan of the $\chi^2$ space is performed. Because the secondary reemission process affects the visible reemitted photons, it has a correlated impact on data sets for all incident wavelengths. At each value of the SQE, a 2-D sweep of VUV QE and absorption length is performed at each wavelength in the ROI. The minimum $\chi^2$ values for each wavelength are summed to determine the global $\chi^2$ value, $\chi^2_{global}(SQE)$, for that choice of SQE. This is shown in Eq. 6 as:

$$\chi^2_{global}(SQE) = \sum_{i=1}^{N} \chi^2_{min,i}. \qquad (6)$$

where N is the number of incident wavelengths considered in the ROI, and $\chi^2_{min,i}$ is the minimum value of the $\chi^2$ found from a 2-D sweep of VUV QE and absorption length for the $i$th incident wavelength. This process is repeated for several different choices of SQE. The SQE value that minimizes $\chi^2_{global}(SQE)$ is taken to be the preferred value. Figure 18 shows the $\chi^2_{global}$ plotted as a function of SQE.

## 7 Results

The results are presented in this section. Section 7.1 presents the measured reemission spectrum for several incident wavelengths. Section 7.2 presents the measured absolute WLSE of TPB films of different thicknesses as a function of incident wavelength. Section 7.3 presents the extracted VUV QE and VUV absorption length of TPB as a function of incident wavelength. Also presented is a value for the SQE.

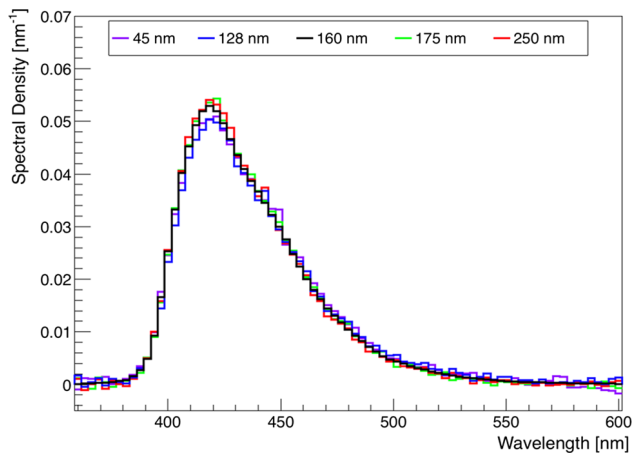

**Fig. 14** The measured reemission spectra of a 1.8 μm TPB film for several incident wavelengths. No dependence of the reemission spectrum of TPB on incident wavelength was observed

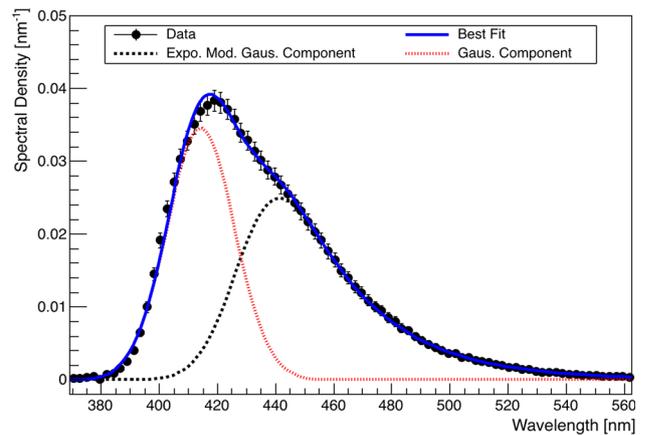

**Fig. 15** The best fit to the TPB reemission spectrum for 160 nm incident light. The reemission spectrum fits well to the weighted sum of a Gaussian (red) and an exponentially modified Gaussian (black). The total fit is given in blue

**Table 3** A table of parameters returned by the best fit to the TPB reemission spectrum is Fig. 15

| Parameter | Value | Uncert. (+/−) |
| --- | --- | --- |
| $A$ | 0.782 | 2.3E-2 |
| $\alpha$ | 3.7E-2 | 5.9E-4 |
| $\sigma_1$ | 15.43 nm | 0.42 nm |
| $\mu_1$ | 418.1 nm | 1.1 nm |
| $\sigma_2$ | 9.72 nm | 0.43 nm |
| $\mu_2$ | 411.2 nm | 0.6 nm |

### 7.1 Reemission spectrum

The visible reemission spectrum was measured for each of several incident wavelengths: 45, 128, 160, 175 and 250 nm. The 45-nm measurement was taken using the SWC and is the brightest peak produced using a HeNe gas mixture. The 128-nm peak measurement was performed using the IWC while the 160-, 175- and 250-nm spectra were taken using the LWC. The 128 and 175 nm wavelengths correspond to the argon and xenon scintillation wavelengths. The area-normalized reemission spectrum for each of these incident wavelengths is presented in Fig. 14.

No dependence of reemission spectrum on incident wavelength was observed. All spectra have a peak near 420 nm and cut off below 400 nm.

In this work, the measured binned spectrum is used in the Monte Carlo model. For the purposes of non-Monte Carlo based modeling, an analytic model is provided here for this spectrum. The reemission spectrum for 160-nm incident light was fit to a weighted sum of a Gaussian and exponentially modified Gaussian:

$$f(\lambda \mid A, \alpha, \sigma_1, \mu_1, \sigma_2, \mu_2) = A \frac{\alpha}{2} e^{\frac{\alpha}{2}(2\mu_1 + \alpha\sigma_1^2 - 2\lambda)} \\ \times \operatorname{erfc}\left(\frac{\mu_1 + \alpha\sigma_1^2 - \lambda}{\sqrt{2}\sigma_1}\right) \\ + (1 - A) \times \frac{1}{\sqrt{2\sigma_2^2\pi}} \\ \times e^{\frac{-(\lambda-\mu_2)^2}{2\sigma_2^2}}. \quad (7)$$

The fit is shown in Fig. 15. The fit was performed using the RooFit package in ROOT [41]. The 160 nm wavelength was chosen because it is the brightest peak of any configuration, which provides the best signal to noise. Because no significant dependence of the reemission spectrum on incident wavelength was observed (Fig. 14), a model built from this fit can be reasonably applied to other incident wavelengths. The fit has a chi-squared value of 0.626. The fit parameters and the associated uncertainties are given in Table 3.

### 7.2 Sample-dependent wavelength shifting efficiency

For comparison to the literature, the measure of sample-dependent WLSE is provided here in the form presented in [16]. The WLSE measured for samples of different thickness are presented in this section.

As discussed in Sect. 6.1, the GAF is determined from simulation and used to convert the observed photon ratio of a sample to the absolute WLSE (Eq. 3). The GAF was found to take an average value of 0.059 with a slight dependence on sample thickness.

Figure 16 presents the measured absolute WLSE efficiency for a representative set of samples: B, C, D, E, and

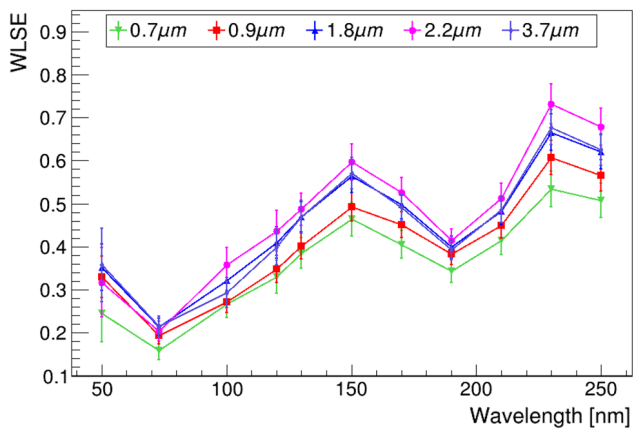

**Fig. 16** Absolute WLSE efficiency for several samples of various thickness

H (as defined in Table 1). In general, thinner samples have a smaller WLSE, up to approximately 2 $\mu$m. Samples between 2- and 3-$\mu$m thick exhibited the largest absolute WLSE. A slight decrease in the WLSE is observed for thickest sample.

It should be emphasized that the WLSE results are dependent on both environmental factors and the exact setup. In this work, the TPB samples were measured at room temperature and in vacuum. For typical LNG applications, the TPB surfaces are often submerged in a cryogenic liquid target. Liquid noble targets have an index of refraction closer to that of acrylic and TPB, so one could expect slightly higher WLSE since there will be fewer reflections at the LNG/TPB and LNG/acrylic interfaces for VUV and reemitted photons. Additionally, one may also see increases in TPB's response when the TPB is at cryogenic temperatures compared to room temperature, as shown in [22].

### 7.2.1 Uncertainties

Several sources of uncertainty were considered. The total uncertainty and its components are plotted as a function of incident wavelength in Fig. 17. Each of the components were added in quadrature to evaluate the total uncertainty, which is provided in Fig. 16.

The photocurrent uncertainty is the RMS in the light, dark, and sample photocurrents propagated through Eq. 2. The "UV Photons" component is the uncertainty in the number of UV photons when folding in the uncertainty in incident light spectrum and photodiode response at the incident wavelengths. Similarly, the "Vis Photons" is the uncertainty in the reemission spectrum of TPB and the corresponding uncertainty of the photodiode's response at the reemission wavelengths.

One photodiode (PD-1) was used for the majority of the measurements. A second photodiode (PD 2) was calibrated relative to a NIST standard and used as a reference to track the calibration of PD-1 over time. The "cross cal" component is the uncertainty in the re-calibration of PD-1 to the PD-2 standard.

When comparing data taken using the LWC and MWC in the region of overlapping measurements (130–150 nm), an offset of 9.8% ± 2% was found. This offset was found to be independent of incident wavelength and was corrected for on all data sets taken using the windowless lamp (MWC and SWC). The configuration dependent offset is likely due to the different illumination profiles on the diffraction grating for the LWC and MWC/SWC. The illumination profile of light incident on the diffraction grating was studied and optimized for the LWC by adjusting the focusing mirror and is thought to be well represented in the Monte Carlo model. This provides reasonable justification to correct the constant offset seen in the MWC/SWC to the LWC.

The uncertainty in the GAF contains two components. The first is the statistical uncertainty of its evaluation in the simulation and is approximately 0.3%. The second and dominant component is the systematic uncertainties of the inputs used in the Monte Carlo model. The systematic uncertainty in dimensions of the setup, as defined in Table 2, was found to be 3.7% which is independent of incident wavelength. To evaluate this systematic, each critical dimension of the setup in turn was changed by ±1$\sigma$ to evaluate the resulting change in the GAF. The uncertainties from each test were assumed to be uncorrelated and were added in quadrature. Other contributions to the systematic uncertainty of the GAF were the surface roughness of the TPB (1.5%), index of refraction of the TPB (3.2%), and scattering length of visible photons in the TPB (2.5%). These uncertainties were obtained by varying these inputs by ±1$\sigma$ from their default values, provided in Sect. 6.2.1, and determining the resulting change in the GAF. The statistical and systematic uncertainties were assumed to be uncorrelated and added in quadrature to yield a total GAF uncertainty of approximately 5.7% which is independent of incident wavelength.

The authors performed an additional systematic check by varying the offset of the TPB sample relative to the photodiode position. This was performed by shimming the TPB sample's seating position in the sample wheel with snap rings. It was found that the change in the observed photon ratio for shimmed configurations was consistent with the change in the GAF given by a simulation of a shimmed setup for 3 shimmed configurations relative to the default configuration. This provides the authors additional confidence that the Monte Carlo simulation is a good representation of the setup.

### 7.3 VUV QE, absorption length, and SQE

The extracted intrinsic VUV QE and absorption length of TPB as a function of incident wavelength and the preferred value of the SQE are presented in this section.

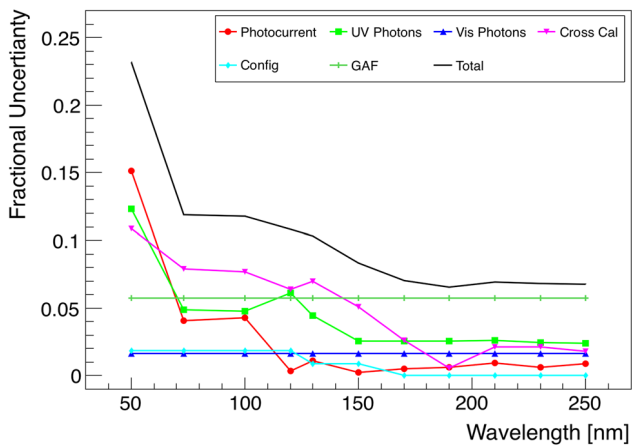

**Fig. 17** Average values for various classes of uncertainties displayed as a function of incident wavelength. The total fractional uncertainty is determined by adding each of the uncertainties in quadrature

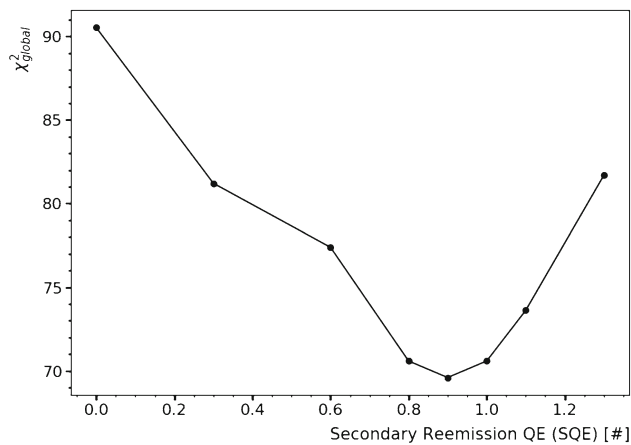

**Fig. 18** The results of $\chi^2_{global}$ plotted as a function of SQE. The minimum at 0.9 is taken to be the preferred value of the SQE. An uncertainty of 0.1 is given by $\pm 1\, \Delta\chi^2_{global}$ from the minimum

Using the process described in Sect. 6.3 and shown in Fig. 18, the SQE was found to have a preferred value of $0.9 \pm 0.1$. The uncertainty in SQE is given by $\pm 1\, \Delta\chi^2_{global}$. The VUV QE and absorption length results presented in this section were determined using SQE equal to 0.9.

The VUV QE values are shown in Fig. 19. The QE extraction was performed using the process described in Sect. 6.3. The VUV QE follows the shape of the WLSE efficiency curves. The QE has local maxima near 230 and 150 nm incident light. There is a general trend toward lower QE for shorter wavelengths.

Recalling the definitions of the VUV QE and WLSE, it is to be expected that the VUV QE is larger than the WLSE. This is because the WLSE efficiency is the intrinsic QE of TPB folded in with the optics of the film, sample and setup. For the scenario where the SQE is less than or equal to 1, the intrinsic QE represents the upper limit for the WLSE which is reduced by the absorption of reemitted photons by the TPB and acrylic.

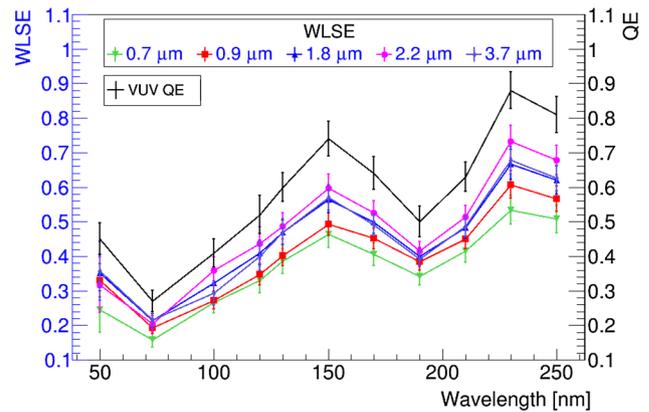

**Fig. 19** The extracted QE of TPB as a function of incident wavelength (right axis). This is plotted alongside WLSE results shown in Fig. 16. These quantities are plotted on different axes because of their fundamentally different definitions

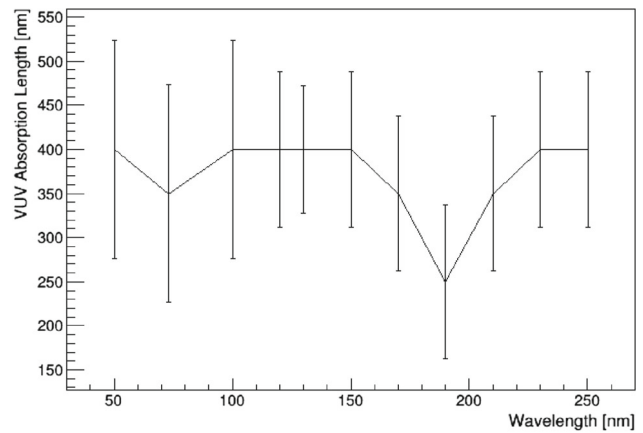

**Fig. 20** The extracted VUV absorption lengths of evaporated thin films of TPB as a function of incident wavelength

The extracted VUV absorption lengths are shown in Fig. 20.

### 7.3.1 Uncertainties

Several sources of uncertainty were considered in evaluating the VUV QE and absorption lengths.

Correlated uncertainties in the VUV QE and absorption lengths are exacted from the 2-D scans at the preferred SQE value (Fig. 13). The uncertainties are evaluated by finding the contour at $+1\Delta\chi^2$ from the minimum and projecting onto the x and y axes [42]. This yields an average uncertainty of 1.8% in VUV QE and 12% in VUV absorption length.

The uncertainty in these quantities due to the SQE was determined by evaluating the change in preferred values of VUV QE and absorption using SQE values $\pm 1\Delta\chi^2_{global}$ from the minimum at 0.9 (Fig. 18). This yields an average uncer-tainty of 4.4% in VUV QE and 12% in VUV absorption length.

In addition, each model input was varied independently within uncertainties, as described in Sect. 6.2.1, to determine the impact on the VUV QE and absorption length result. The systematic uncertainty from the geometry's dimensions was 4% for the QE. The uncertainty from TPB's index of refraction was 1.8% for the VUV QE and 12% for the absorption length. The uncertainty from the visible scattering length was 3.3% for the VUV QE. No impact on the absorption length results was observed when changing the geometry's dimensions, TPB index of refraction, or visible scattering length.

Adding these uncertainties in quadrature yields an average uncertainty of approximately 8% in VUV QE and 26% in VUV absorption length. The uncertainties shown in Figs. 19 and 20 include all contributions.

## 8 Discussion

The key results of this work are centered around extracting important microphysical parameters critical to modeling TPB thin films in detector simulations. In particular, measurements of the intrinsic QE and absorption lengths of room temperature TPB thin films are presented as a function of incident wavelength in the VUV regime. Also presented is a measurement of the SQE – the average efficiency of secondary reemission of visible photons in the regime where the TPB absorption and reemission spectra overlap.

In an effort to directly compare with results presented in [16], the sample-dependent WLSE of several samples of different thickness and reemission spectra are also presented as a function of incident wavelength. These results cover the spectral range studied by [16] with improved precision, while extending the measurements down to 50 nm.

To the best of our knowledge, the extraction of the SQE and intrinsic VUV QE and absorption lengths as a function of incident wavelength for evaporated TPB thin films, without reference to other materials, is a new result. These intrinsic properties are more broadly applicable than the blackbox WLSE, because they depend on the properties of the material rather than specifics of the sample. For the purposes of modeling TPB, the intrinsic VUV QE, absorption length, and SQE are critical inputs to current and future liquid noble gas experiments and are now well constrained.

The WLSE result presented in this work is in tension with those presented in [16]. The authors believe the differences can be attributed to unaccounted for systematic uncertainties related to photodiode calibration and the setup optics in [16], and to differences in the Monte Carlo model used to evaluate the GAF. The details of the differences are discussed in Sect. 8.1.

Previous measurements of the WLSE of TPB were performed relative to Sodium Salicylate [20]. Combining the Sodium Salicylate reference [26] with the relative measurements suggests the QE of TPB should be larger than that determined in this work. The authors believe that this difference is being driven by the optical model of the TPB, which may not be effectively represented in the Sodium Salicylate reference. Also presented in [20] is the relative response of samples as a function of thickness. Data and Monte Carlo predictions from this work are compared in a relative fashion to the data in [20] as an independent check. This is discussed in Sect. 8.2.

The reemission spectrum result is consistent with the literature and has confirmed that the reemission spectrum of room temperature TPB thin films remains constant for incident wavelengths as low as 45 nm.

### 8.1 Comparison with previous measurement

This work builds on much of the original work in [16], and several items of equipment were shared between the two efforts. The authors of this article worked closely with the authors of [16], one of whom is an author of this work.

Differences exist between the TPB efficiency measurements presented here and what is presented in [16], as shown in Fig. 22. Several factors can account for the differences in the measurements. These are described in Sects. 8.1.1–8.1.4. Following many detailed systematic checks, as described, there is confidence that the differences are understood.

#### 8.1.1 Photodiode calibration

The authors believe the largest differences between this work and what is presented in [16] arise from issues with photodiode calibrations. The photodiode used in [16] was calibrated in 2008 by NIST and had the response given by the solid curve in Fig. 21 [43]. Before the start of this work, the same photodiode was recalibrated at NIST and followed the response curve given by the dotted curve [44]. A large change in the photodiode's response occurred between 2008 and 2014, especially in the range of 120–150 nm. The previous work was published in 2011 [16].

The photodiode's response degradation was likely due to UV damage from prolonged exposure and the build up of an oxidation layer on the bare face of the silicon photodiode surface [44,45]. A build up of an oxidation layer can drastically reduce the response of the photodiode for wavelengths below 150 nm. Interestingly, an oxide layer may also increase the response at longer wavelengths because the layer creates an anti-reflective layer (interactions inside of the oxide film). An oxide layer effectively leaves the response at the visible wavelengths unchanged. This explanation is consistent with what is seen in Fig. 21.

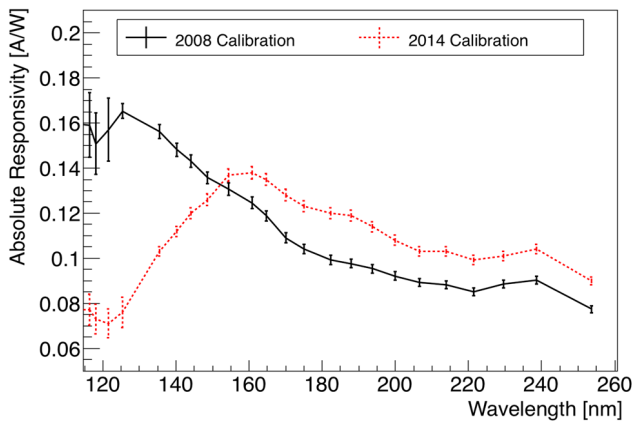

**Fig. 21** A comparison comparison of the NIST calibrations of the photodiode used in [16] in 2008 (solid) and in 2014 (dotted). The 2008 curve was used as the photodiode response function in [16]. There is a substantial difference between the 2008 and 2014 response functions, suggesting that significant degradation in the photodiode's response occurred between 2008 and 2014

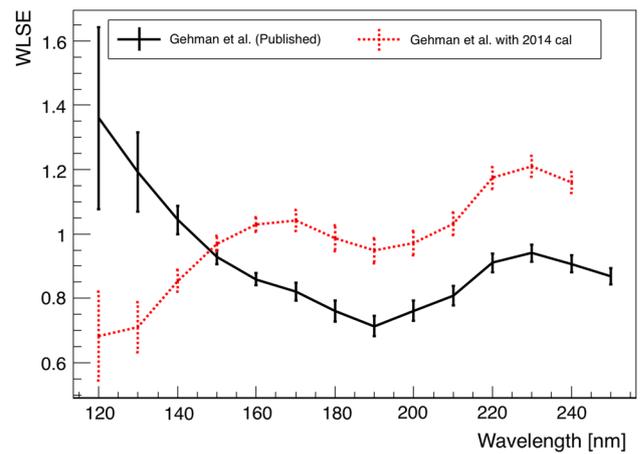

**Fig. 22** A comparison of the published results in [16] (solid line) to a reanalysis of the raw data from [16] using the 2014 calibration (dotted line). When the 2014 calibration is applied, the result from [16] has a similar shape as the results from this work. The scale offset is explained in Sect. 8.1.2

Because the response of the photodiode used in the 2011 publication changed by a large amount, two new photodiodes were purchased in 2016 after the start of this work. NIST calibrated one of these photodiodes (PD 1) (Fig. 5) in February of 2016. The second photodiode (PD 2) was calibrated relative to the NIST supplied calibration of PD 1 by the authors. PD 1 was used for the bulk of our measurements in this work while PD 2 was used to track the relative response of PD 1 over time.

With access to the raw data from the 2011 publication, another analysis was performed using the 2014 calibration of the photodiode used in [16]. Figure 22 shows the results of this reanalysis as a dotted line, along with the published 2011 result from [16] as a solid line. For short wavelengths (120–150 nm), an unaccounted degradation in the photodiode's response would lead to an underestimation of the number of UV photons incident on the sample, thus increasing the photon ratio and WLSE at those wavelengths. The opposite is true for the longer wavelengths where an increase in the photodiode's response was observed (150–250 nm).

Interestingly, the shape of the reanalyzed raw data from the 2011 result using the 2014 calibration curve agrees fairly well with the results presented in this work, though there is a normalization offset which will be explained in Sect. 8.1.2. This suggests that the bulk of the Gehman et al. photodiode's response degradation occurred between the 2008 calibration and the 2011 publication and that the published result does not account for changes in the photodiode's response.

### 8.1.2 GAF determination

The GAF, described in Sect. 6.1, determines a scale factor to evaluate a sample's WLSE from the photon ratio (otherwise known as the forward efficiency in [16]). As may be expected, the GAF is observed to have a strong dependence on the distance of the photodiode from the sample. This suggests careful measurements of the apparatus dimensions are required and careful treatment of these uncertainties is critical. This systematic uncertainty has been included in our result and is the dominant uncertainty contribution for several wavelengths. The work presented in [16] also used a Monte Carlo simulation to determine the GAF. The model was less detailed than that used in this work and the systematic uncertainty on the GAF was assumed to be zero. Additionally, the TPB model in [16] was treated as an infinitesimally thin film which did not account for important TPB optical effects, such as the scattering length and secondary reemission of visible light. The authors believe the simplified Monte Carlo model in [16] underestimated the value of the GAF which led to systematically higher WLSE results.

As a crosscheck, the setup and sample used in [16] was modeled using the Monte Carlo described in this work, to determine the impact on the WLSE of using the more sophisticated model to evaluate the GAF. Differences between the setup used in [16] and this work include a thicker acrylic substrate and the distance between the TPB surface and photodiode. Both of these differences were accounted for in the evaluation of a GAF for the apparatus used in [16]. Applying this newly-evaluated GAF to the results in Fig. 22 yields Fig. 23. The shaded region spanning the dashed and dotted lines illustrates the uncertainty in the photodiode calibration used in [16] at the time of the 2011 measurement. For comparison, the WLSE from this work for a sample of comparable thickness is plotted as a solid line. This result is consistent with the GAF-corrected results from [16] to within the limits

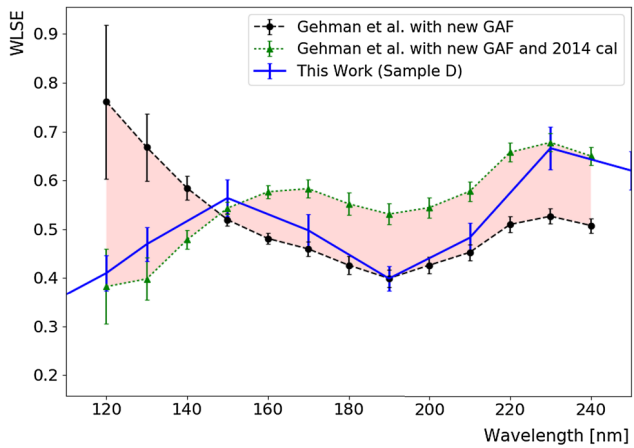

**Fig. 23** Impact of re-analyzing data from [16] using the Monte Carlo model described in this work. The solid line shows the WLSE of a sample of comparable thickness to that studied in [16]. The shaded region spanning the dashed and dotted lines illustrate the uncertainty in the photodiode calibration used in [16] at the time of the 2011 measurement

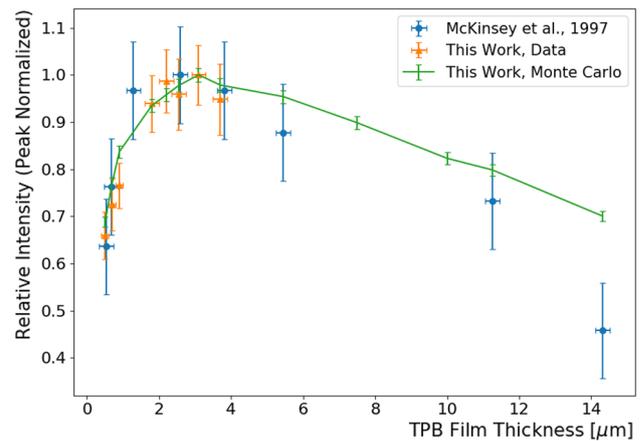

**Fig. 24** A comparison of the relative response of samples as a function of TPB film thickness for evaporated TPB thin film data in [20], data from this work, and Monte Carlo model predictions with the model used in this work

driven by the uncertainty in that work's photodiode calibration.

These differences in evaluating the GAF and its uncertainty contribute to differences in scale between the WLSE measurements presented in this work and those presented in [16].

### 8.1.3 Background subtraction and beam intensity

This work used much of the same hardware as was used in [16]. It was determined that the original focusing mirror and Al + MgF$_2$ diffraction grating required replacement because of discoloration on the optical surfaces likely resulting from years of use. It was observed that the level of background light was elevated and the intensity of light at short wavelengths was attenuated when using the discolored optical elements. When the damaged optical elements were replaced at the start of this work, the amount of background light visible by the spectrometer was substantially reduced and the VUV light output at wavelengths between 120 and 250 nm was improved by between 20 and 50% across the wavelength range of interest.

A brighter light source leads to smaller statistical errors in the ratio of measured dark and background subtracted photocurrents given in Eq. (3). This is a dominant reason why the statistical error in the WLSE measurements of this work are smaller at most wavelengths than in [16]. Higher statistics resulting from a more intense incident UV beam also result in the cleaner, more stable reemission spectra presented in this work (Fig. 14).

It is also important to note that in this work, a background subtraction is performed in addition to a dark current subtraction for wavelengths below 150 nm (Eq. (1)). Because any measurement of reemitted light from a TPB sample is actually a sum of reemitted light plus any background light, a separate measurement is required to determine the magnitude of the background light, as described in Sect. 5.1. This correction was not included in [16] – only a dark current subtraction was performed. Background levels at short wavelengths (below 135 nm) were observed to be on the order of the reemitted signal when using the discolored optics inherited from the setup used in [16]. It is possible that elevated background levels may have been unaccounted for in their TPB signal measurements.

### 8.1.4 Overlapping measurements

This work was able to leverage two distinct light sources to make multiple measurements of the total wavelength shifting efficiency in the range of 130–150 nm. Referred to as the overlap region, the ability to measure the WLSE in this region using multiple configurations and gases provides a nice cross-check of configuration dependent systematic uncertainties. All measurements from the various configurations in this overlap region agree well, providing confidence that there is reasonable control of configuration-dependent systematic uncertainties.

## 8.2 Previous QE reference

As an additional check of the thickness dependence of our data and model, results were compared to the relative response of samples of various thickness in [20]. Figure 24 shows the relative response of samples as a function of film thickness where the sample with the largest response is normalized to 1 for 74 nm incident light. The data reported in [20] is a function of coating thickness (mg/cm$^2$) and was converted to film thickness in micrometers using the inferred evaporated film density of 1.46 g/cm$^3$ from [16] and the assumption that the evaporated films in [20] were uniform.

As seen in Fig. 24, the peak-normalized data from [20], data from this work, and the Monte Carlo model used in this work are in fairly good agreement. This comparison provides a crosscheck of the thickness-dependent behavior predicted by the micro-physically motivated Monte Carlo model developed in this work, demonstrating that this model is consistent with independent data sets, and can be used to extrapolate to film thicknesses greater than those used in this work.

It should be emphasized that all of the values used in the Monte Carlo simulation in this work, except for the VUV QE, absorption length and SQE, were set by measurement (remission spectrum) or values in the literature. Given the highly constrained nature of this model, the fact that the model accurately predicts the dependence of photon ratio on sample thickness (Figs. 12, 24) suggests that the Monte Carlo model used in this work is a good representation of the optical properties of TPB.

## 9 Conclusions

This paper presents the first measurement of certain microphysical parameters of TPB wavelength-shifting films. This leads to development of a complete model of the optical properties of vacuum-deposited TPB thin films at room temperature, independent of optical effects in the setup and the samples themselves, which is thus broadly applicable to other detectors. Results include the first measurement of the intrinsic QE and absorption lengths in the VUV regime as a function of wavelength. Also presented is a measurement of the SQE – the average efficiency of secondary reemission of visible photons in the regime where the TPB absorption and reemission spectra overlap.

These results enable high-precision modeling of VUV light detection in liquid noble gas experiments, such as those probing cutting-edge neutrino and dark matter physics.

**Acknowledgements** This work was supported in part by the Office of High Energy Physics of the U.S. Department of Energy under contract DE-AC02-05CH11231, and in part by the Physics Department at the University of California, Berkeley. Work at the Molecular Foundry was supported by the Office of Science, Office of Basic Energy Sciences, of the U.S. Department of Energy under Contract No. DE-AC02-05CH11231. The Ocean Optics spectrometer, McPherson monochromator, deuterium light source, lamp power supply, monochromator motor controller, and two photodiodes are kindly on loan from Dr. K. Rielage and Los Alamos National Laboratory. The authors thank the MiniCLEAN collaboration for their permission to use MiniCLEAN RAT and Dr. Thomas Caldwell for his input on modeling TPB in MiniCLEAN RAT. The authors thank the Dr. Daniel McKinsey group at University of California at Berkeley for the use of their thermal evaporator and for valuable discussion.

## References

1. V. Chepel, H. Araujo, JINST **8**, R04001 (2013)
2. S. Amerio et al. (ICARUS Collaboration), Nucl. Instrum. Methods A **527**, 329–410 (2004)
3. R. Brunetti et al. (WARP Collaboration), New Astron. Rev **49**, 265 (2005)
4. C. Amsler et al. (ArDM Collaboration), Acta Phys. Polon. B **41**, 1441–1446 (2010)
5. P. Agnes et al. (DarkSide Collaboration), Phys. Lett. B **743**, 456–466 (2015)
6. M.G. Boulay (DEAP Collaboration), J. Phys. Conf. Ser. **375**, (2012)
7. K. Rielage et al. (MINICLEAN Collaboration), Proceedings, 13th International Conference on Topics in Astroparticle and Underground Physics (TAUP 2013), Phys. Proc. **61**, 144 (2015)
8. R. Acciarri et al. (MicroBooNE Collaboration), JINST **12**, P02017 (2017)
9. R. Acciarri et al. (DUNE Collaboration), (2016). arXiv:1512.06148
10. D.Y. Akimov et al. (ZEPLIN Collaboration), Astropart. Phys. **27**, 46–60 (2007)
11. E. Aprile et al. (XENON Collaboration), Astropart. Phys. **35**, 573–590 (2012)
12. D.S. Akerib et al. (LUX Collaboration), Nucl. Instrum. Methods A **704**, 111–126 (2013)
13. D.S. Akerib et al. (LZ Collaboration), arXiv:1509.02910 [physics.ins-det]
14. K. Abe et al. (XMASS Collaboration), Nucl. Instrum. Methods A **716**, 78–85 (2013)
15. J. Aalbers et al. (DARWIN Collaboration), JCAP **1611**, 17 (2016)
16. V.M. Gehman et al., Nucl. Instrum. Methods A **654**, 116–121 (2011)
17. J.A.R. Samson, Wiley, Amsterdam (1967)
18. W.M. Burton et al., Appl. Opt. **12**, 87 (1973)
19. S.E. Wallace-Williams et al., J. Phys. Chem. **98**, 60–67 (1994)
20. D. McKinsey et al., Nucl. Instrum. Methods B **132**, 351–358 (1997)
21. B. Jones et al., JINST **18**, P01013 (2013)
22. R. Francini et al., JINST **8**, P09006 (2013)
23. D. Stolp et al., JINST **11**, C03025 (2016)
24. S.P. Regan et al., Appl. Opt. **33**, 3595 (1994)
25. R. Allison et al., Opt. Soc. Am. **54**, 747–750 (1964)
26. E.C. Bruner Jr., Opt. Soc. Am. A **59**, 204 (1969)
27. http://mcphersoninc.com/pdf/634.pdf. Accessed Aug 2017
28. http://www.mcphersoninc.com/pdf/234302.pdf. Accessed Aug 2017
29. http://optodiode.com/pdf/AXUV100G.pdf. Accessed Aug 2017
30. L.R. Canfield et al., Metrologia **35**, 329 (1998)
31. E.M. Gullikson et al., J. Electr. Spectrosc. Relat Phenomena **80**, 313 (1996)
32. R. Vest, NIST Photodiode Calibration Report (2016)
33. Polymer Plastics Company, LC. UVT Acrylic Ultraviolet Transmitting Sheet. http://www.polymerplastics.com/transparents_uvta_sheet.shtml. Accessed Aug 2017
34. R. Brun, F. Rademakers, Nucl. Instrum. Methods A **389**, 81–86 (1997)

35. M. Aoki, Y. Iwashita, M. Kuze, T. Bolton, Nucl. Phys. B (Proc. Suppl.) **149**, 166 (2005)
36. S. Andringa et al. (SNO+ Collaboration), Adv. High Energy Phys. **2016**, 6194250 (2015)
37. M. Kuzniak. DEAP/CLEAN internal documentation. Document ID cl0911010. Personal Communication. 10 November 2009
38. D. Bower, *An Introduction to Polymer Physics* (Cambridge University Press, New York, 2002)
39. H. Huber et al., Can. J. Chem. **42**, 2065 (1964)
40. http://geant4.slac.stanford.edu/UsersWorkshop/PDF/Peter/OpticalPhoton.pdf. Accessed Oct 2017
41. W. Verkerke, D. P. Kirkby, The RooFit toolkit for data modeling. In Proceedings of 2003 Conference for Computing in High-Energy and Nuclear Physics (CHEP 03), La Jolla, California, 24–28 March 2003, eConf C0303241 (2003) MOLT007. arXiv:physics/0306116
42. W.H. Press et al., *Numerical Recipes in C: The Art of Scientific Computing* (Cambridge University Press, New York, 1992)
43. R. Vest, Report of calibration provided by National Institute of Standards and Technology. Internal documentation (2008)
44. R. Vest, Report of calibration provided by National Institute of Standards and Technology. Internal documentation (2014)
45. C. S. Tarrio et al., Proc. SPIE, **7271**, (2009)